\pdfoutput=1
\documentclass[conference]{IEEEtran}
\IEEEoverridecommandlockouts
\usepackage{subcaption}
\usepackage{cite}
\usepackage{amsmath,amssymb,amsfonts}
\usepackage{algorithmic}
\usepackage{graphicx}
\usepackage{textcomp}
\usepackage{xcolor}

\def\BibTeX{{\rm B\kern-.05em{\sc i\kern-.025em b}\kern-.08em
    T\kern-.1667em\lower.7ex\hbox{E}\kern-.125emX}}
\begin{document}

\title{Improving Power Quality and Power Sharing in Unbalanced Multi-Microgrids Using Energy Storage System\\
}
\author{\IEEEauthorblockN{Mehmet Emin Akdogan, and Sara Ahmed}
\IEEEauthorblockA{\textit{Electrical and Computer Engineering} \\
\textit{The University of Texas at San Antonio}\\
San Antonio, Texas \\
m.eminakdogan@gmail.com, sara.ahmed@utsa.edu}
}

\maketitle
\begin{abstract}
The integration of single-phase distributed generations (DG) and unbalanced loads to three-phase DGs brings challenges to the operating system of the  microgrids (MGs). These challenges include unbalanced voltage and unequal power flow at the point of common coupling (PCC) of the MGs. The aim of this paper is to develop an Energy Storage System (ESS) with multi-function control for islanded multi-microgrids (MMG) consisting of single and three PV-DGs to enhance its power quality and power sharing of the MMGs. The proposed multi-function control consists of a reactive power compensation (RPC), a PCC power balance regulator (PBR) and a reactive power sharing algorithm (RPSA). The RPC and the PBR improve the three-phase PCC voltage and power unbalance of the MMG and the RPSA regulates the reactive power sharing among the MMGs. The proposed strategy is experimentally validated on the Opal-RT OP5600 real-time simulator. The voltage unbalance factor (VUF) at the PCC is reduced from 4.3 percent to 0.03 percent, while the three phase and the single phase reactive powers are shared proportionally between different rated DGs.

\end{abstract}


\section{Introduction}
With the development of DG technologies, distributed photovoltaic (PV) energy sources have gained widespread attention recently. This is due to its advantages including emission reduction and improved energy efficiency \cite{MMG,05}. DG-PVs are connected to microgrids (MGs) and operate locally through power electronics interface converters. A microgrid is a local grid that integrates multiple parallel DG units, energy storage, and backup generators to improve the reliability of the power system operation\cite{02,03,05,06,KHAN2021,MMG,akdogan}.

The number of residential rooftop single phase PV-DGs in three phase MGs have increased significantly in the past decades \cite{al,mypaper,MMG}. This high penetration of single-phase PV-DGs results in power quality issues such as voltage unbalance and DGs proper power sharing\cite{MMG,HMG1,sequence,Multiobjective, RPCSP1,RPCSP2,RPCSP3, ESS1,ESS2,Droop1,ESShybrid, apf,af,HMG2,Droop1,Droop2,ESSdroop1,ESSdroop2,ESS-singlephase}. In addition, the existence of a single-phase load of the single-phase microgrid also impacts MMGs and produces unequal power in three phase MMGs.

Various DG inverter voltage regulation strategies have been proposed in literature to deal with voltage quality issues in MMGs  and low voltage distribution networks (LVDNs) \cite{MMG,HMG1,sequence,Multiobjective, RPCSP1,RPCSP2,RPCSP3, ESS1,ESS2,Droop1,ESShybrid, apf,af,HMG2,Droop1,Droop2,ESSdroop1,ESSdroop2,ESS-singlephase}. A power management strategy based on a hierarchical event-based distributed control method \cite{HMG1} has been proposed to enhance the voltage unbalance at PCC and DG terminals for hybrid AC/DC MGs.  
Positive, negative, and zero sequence-based current controllers with reactive power compensation method in \cite{sequence} are presented to control unbalanced voltage for PV-DGs connected unbalanced distribution network. In addition, the optimal multiobjective control algorithm based on a master-slave architecture is implemented in \cite{Multiobjective} to mitigate the voltage unbalance, optimally regulate power flow, and also minimize the reactive power circulation using randomly connected single-phase DERs without requiring previous knowledge of the MG parameters. 

Some studies have used single-phase DG inverters by compensating reactive power to regulate the unbalanced voltage in MMGs\cite{RPCSP1,RPCSP2,RPCSP3}. The authors of \cite{RPCSP1} proposed a consensus algorithm that coordinates the single-phase PV inverters using the reactive power capacity of the inverters for voltage regulation. A flexible unbalance compensation strategy is introduced in \cite{RPCSP2} to mitigate voltage unbalance system and share the power demand among DGs proportionally based on their available power. However, the reactive power control using the single-phase DG inverters may reduces inverters’ lifetime without any financial benefit for residential PV customers\cite{ESS1,ESS2}. 
\begin{figure*}[htbp]
\centering
\includegraphics[width=\linewidth]{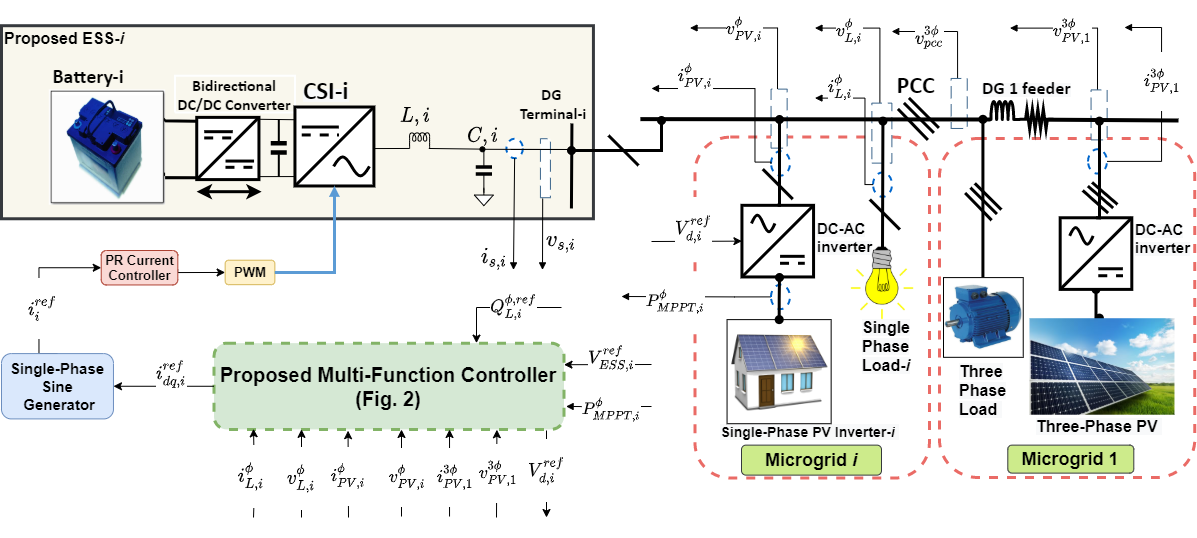}
\caption{Simplified block diagram of the proposed ESS control connected to unbalanced multimicrogrid system.}
\label{Simplified}
\vspace{-5pt}
\end{figure*}

In addition, another solution is the connection of a single-phase ESS to a single-phase PV-DG in parallel \cite{ESS1,ESS2}. Distributed battery energy storage (BES) systems  are used to mitigate the voltage unbalance in the network \cite{ESS2}.  \cite{ESS1} proposes a consensus-based distributed control strategy for BESs systems connected to rooftop PV systems for voltage regulation.  The authors in \cite{ESS1,ESS2}  only developed the voltage regulation methods for LVDNs and not for MMGs. In addition, a hybrid storage topology in the MMG provides active balancing for mitigation of the phase unbalances \cite{ESShybrid}.

The integration of single-phase PV systems and unbalanced loads results in unequal power flow in the three-phases at the PCC terminal and leads to unbalanced conditions. A hierarchical coordinated strategy including primary and secondary controls to regulate the voltage imbalance and the three-phase power unbalance at the terminal of the PCC for the islanded MMG \cite{MMG}. However, using only secondary control cannot solve the unbalanced power issue at the PCC. 
Another problem is the reactive power sharing in three-phase and single-phase PV systems. Droop control methods have been proposed for primary control of the DG inverters to regulate the active and reactive power sharing and compensate reactive power for voltage regulation\cite{Droop1}. However, a single-phase droop control in \cite{Droop1} applied to systems with unbalanced single-phase loads cannot share the single-phase and the three-phase reactive power appropriately for different rated MGs.

A multi-function control to provide both reactive power compensation and sharing by regulating single phase ESSs is addressed in only a few references. This paper uses ESS with improved multi-function control to mitigate the unbalanced voltage and power and solve the reactive power sharing problems caused by the integration of single-phase PV systems in three-phase MGs. The reactive power compensation (RPC) sets the reference of the active power of the ESS for voltage quality improvement. The power balance regulator (PBR) is presented to balance active and reactive powers of the three-phase PV systems in the unbalanced MGs and the reactive power sharing algorithm (RPSA) is used to achieve reactive power sharing among the three-phase and the single-phase PV inverters. The main contributions of the paper is as follows: 

\begin{itemize}
\item 	A new multi-function power quality control strategy based on the incremental algorithm is developed for reactive power compensation and sharing, and balancing three-phase power at the PCC terminal. 
\item 	A basic incremental algorithm (IA) is applied to each controller of the ESS (RPC, PBR, RPSA) to  reach a desired set point by incrementally changing the output signal. This algorithm improves the system performance accuracy and dynamic characteristics when compared to conventional PID controller.
\item The design of the proposed control avoids complex calculations and requires only line  voltage  measurements.
\end{itemize}

The remaining parts of the paper are presented as follows: in Section II, the system structure and proposed control scheme, including IA, RPC, PBR and RPSA are presented. Section III is dedicated to present the real time simulation results and finally, this paper is concluded in Section IV.

\begin{figure*}[htbp]
\centering
\includegraphics[width=0.85\linewidth]{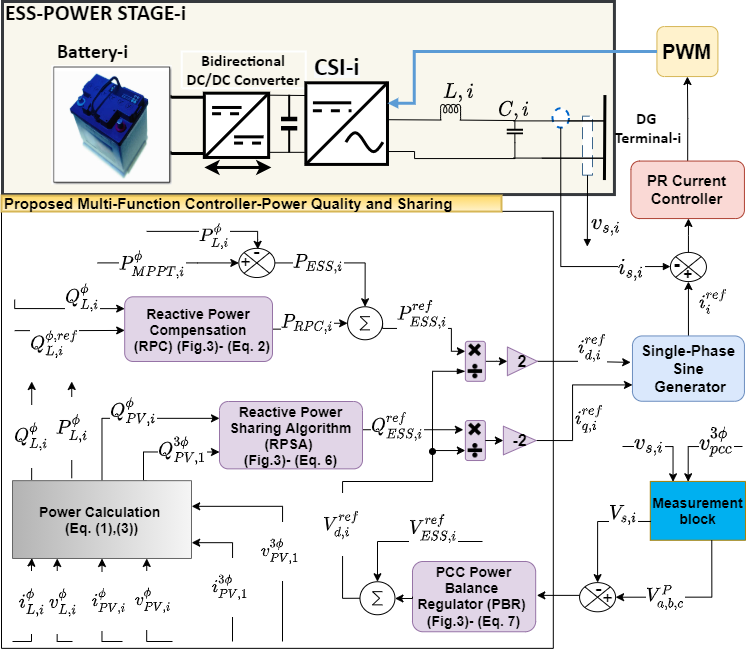}
\caption{Detailed block diagram of proposed control scheme for the ith ESS.}
\label{Detailed}
\vspace{-5pt}
\end{figure*}

\section{the system structure and  proposed multi-function controller scheme}
Fig.~\ref{Simplified} shows the single-line diagram of the MMG model under study. The system is composed of two types of PV-MGs; microgrid 1 is a three-phase PV-MG and microgrid  $i$ is a single-phase PV-MG connected to one phase of the three-phase PV-MG. The proposed ESS is connected in parallel to the single-phase PV system and to a single-phase load as shown in Fig.~\ref{Simplified}. The power stage of ESS $i$ and its proposed control scheme block diagram are demonstrated in Fig.~\ref{Detailed}. Each power stage comprises of a battery storage unit connected to a bidirectional dc/dc converter feeding a current source inverter (CSI).


The integration of the single-phase and the three-phase PV systems changes the active and reactive power seen at the single-phase load. 
This proposed multi-function controller adjusts the single-phase load active power ($P_{L,i}^{\phi}$) and reactive power ($Q_{L,i}^{\phi}$) by generating a proper output current ($i_{i}^{ref}$) of the ESS inverter to improve power quality and sharing at the three-phase PCC terminal in Fig.~\ref{Detailed}. The detailed control loops will be discussed in this section. The proposed multi-function control strategy is as follows:

\subsection{Incremental Algorithm (IA)}
\begin{figure}[!t]
\centerline{\includegraphics[width=1.\linewidth]{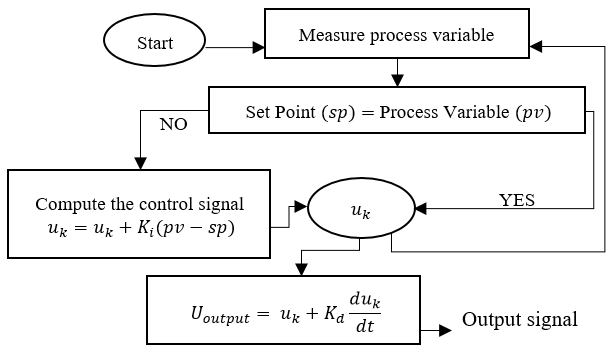}}
\caption{ The flowchart of incremental algorithm for the ith ESS.}
\label{AI}
\end{figure}

The flowchart of the IA is shown in Fig.~\ref{AI}. A basic algorithm minimize an error value as the difference between a desired set point ($sp$) and a measured process variable ($pv$). The algorithm with a derivative term is implemented to each controller in the ESS to reach $sp$ by incrementally changing the control signal (${u}_k$). Then ${u}_k$ and derivative of ${u}_k$ added together to generate the proper output signal. 

Compare to conventional PID controller, derivative term is applied to ${u}_k$ (not to the error value) to reduce the overshoot and improve stability. In addition, integrator windup for the PID controller is an issue that has to be considered. The windup is the process of accumulating of an error sum when the process variable is approaching the set point and may cause a significant amount of overshoot and inefficient control. However, for the incremental form, there is no summation which means no error accumulation or integrator windup as the IA inherently contains an anti-windup \cite{PID,PID2}. The approach is adapted in all the multi-function controller loops; RPC, RPSA and PBR. 
\vspace{-6pt}
\subsection{Reactive Power Compensation (RPC)} The proposed RPC compensates for the unbalanced voltage at the PCC. 
Average reactive power of a single-phase load ($Q_{L,i}^{\phi}$) is calculated using the instantaneous power passed through a first-order low pass filter (LPF).  $Q_{L,i}^{\phi}$ is shown in \eqref{ql}:
\begin{equation}
Q_{L,i}^{\phi}=(w_{LPF}/(s + w_{LPF})).(\frac{-1}{2}(v_{Ld,i}^{\phi}.i_{Lq,i}^{\phi})).
\label{ql}
\end{equation}
where $v_{Lq,i}^{\phi}$, $i_{Ld,i}^{\phi}$ are the voltage and current of the single-phase load in dq frame respectively\cite{daxis}. Then $Q_{L,i}^{\phi}$ is compared with a reference reactive power of the single-phase load ($Q_{L,i}^{\phi,ref}$) in the RPC to generate a reactive power compensation reference ($P_{RPC,i}$) as shown in the following equation: 
\begin{align} 
P_{RPC}=(Q_{L,i}^{\phi ,ref}-Q_{L,i}^{\phi}).{U}_\mathit{output}.
\label{RPC}
\end{align}
In addition, active power of a single-phase load ($P_{L,i}^{\phi}$) is calculated and passed through a LPF as shown in \eqref{Pl1}:
\begin{equation}
P_{L,i}^{\phi}=(w_{LPF,i}/(s + w_{LPF,i})).(\frac{1}{2}(v_{Ld,i}^{\phi}.i_{Ld,i}^{\phi})).
\label{Pl1}
\end{equation}
The charge/discharge power of the ESS is calculated as follows: 
\begin{equation}
P_{ESS,i}=P_{MPPT,i}^{\phi}-P_{L,i}^{\phi}.
\label{PESS}
\end{equation}
where $P_{MPPT,i}^{\phi}$ is the maximum power generated by the $ith$ single-phase PV array. The reference active power of the ESS for improving voltage quality is defined as:
\begin{equation}
P_{ESS,i}^{ref}=P_{ESS,i}+P_{RPC,i}.				
\label{Pl}
\end{equation}
\begin{figure}
\centering
\includegraphics[width=0.8\linewidth]{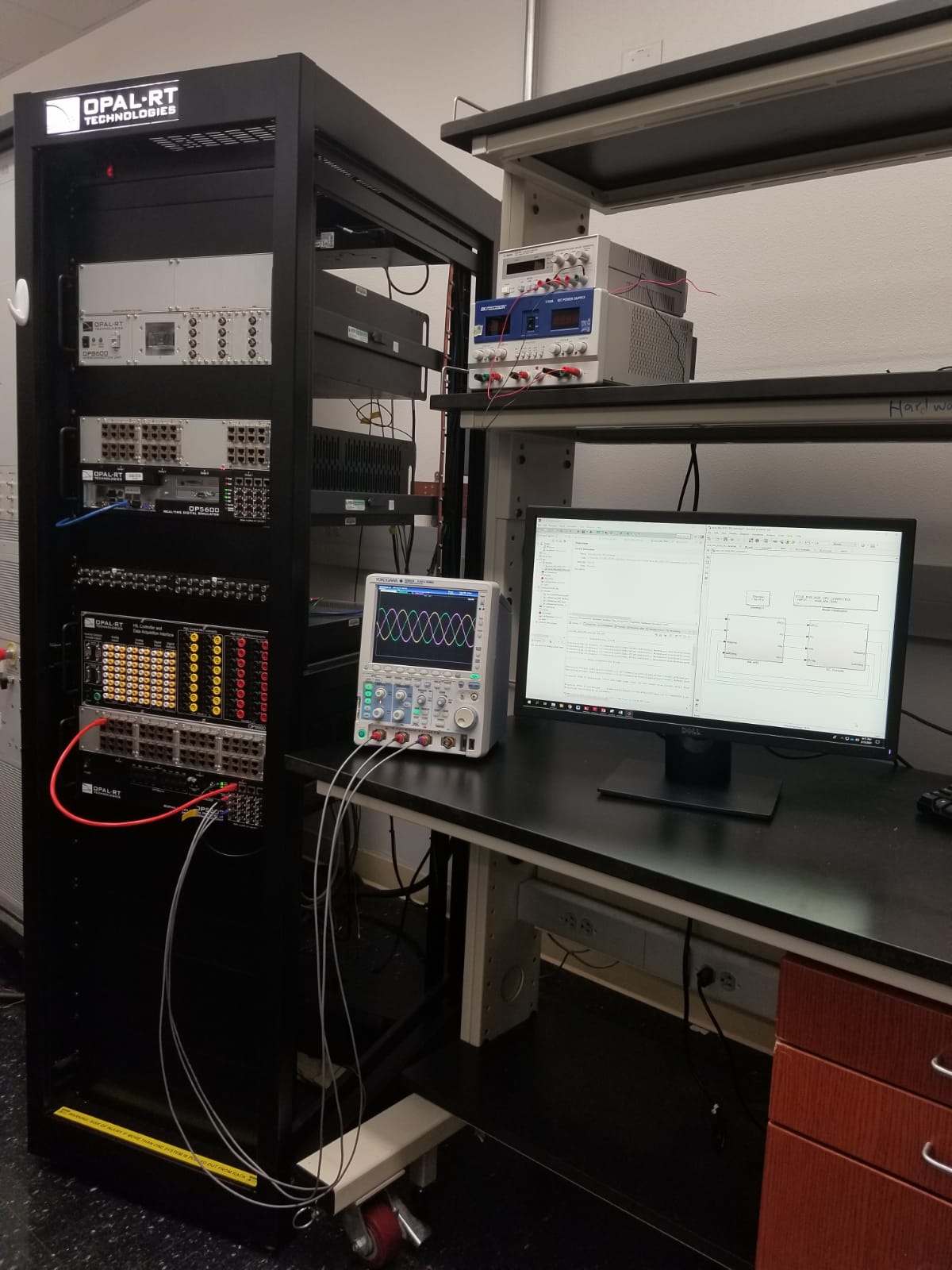}
\caption{Opal-RT OP5600 real-time simulator.}
\label{opal}
\vspace{-10pt}
\end{figure}

\subsection{Reactive Power Sharing Algorithm (RPSA)} Compensating $Q_{L,i}^{\phi}$ changes reactive power of the single-phase PV inverter ($Q_{PV,i}^{\phi}$) and affects adversely reactive power sharing between the single-phase and the three phase inverters. 
The same equation in (1) is used to calculate $Q_{PV,i}^{\phi}$ and the calculation of $Q_{PV,1}^{3\phi}$ is discussed in \cite{mypaper}. 

The reference for the reactive power sharing algorithm ($Q_{PV,i}^{\phi,ref}$) is the reactive power of the three-phase PV inverter ($Q_{PV,i}^{3\phi}$) 
divided by the ratio of the three-phase PV–DG rated power capacity($P_{PV,i}^{3\phi}$) and the single-phase PV–DG rated power capacity ($P_{PV,i}^{\phi}$). Note that the three-phase inverter is 4 times the power capacity of the single-phase PV inverters under study. The reference reactive power of the ESS ($Q_{ESS,i}^{ref}$) in \eqref{RSPA1} is produced to the share reactive power proportionally between PV inverters
\vspace{1.5pt}
\begin{align} 
\begin{split}
 Q_{PV,i}^{\phi,ref}= Q_{PV,i}^{3\phi}/(P_{PV,i}^{3\phi}/P_{PV,i}^{\phi})
\\Q_{ESS,i}^{ref}= (Q_{PV,i}^{\phi,ref}-Q_{PV,i}^{\phi}).{U}_\mathit{output}.
\end{split}
\label{RSPA1}
\end{align}
\vspace{-6pt}
\subsection{PCC Power Balance Regulator (PBR)} In unbalanced MMG, the phase power for phases A, B, and C of the three-phases at the PCC terminal is still not balanced because of injecting active power from the three-phase PV inverter to the single-phase inverter. This controller is applied to mitigate the unbalanced voltage and regulate the three-phase power at the PCC terminal. The magnitude ($V_{s,i}$) of the ESS output voltage ($v_{s,i}$) and the voltage magnitudes of the phases ($V_{a}^{P}$, $V_{b}^{P}$, $V_{c}^{P}$) from the three-phase PCC voltage ($v_{pcc}^{3\phi}$) are measured by phase locked loop (PLL) in the measurement block. The reference output voltage of the ESS $i$ or single-phase PV inverter ($V_{d,i}^{ref}$) is expressed as: 
\begin{equation}
V_{d,i}^{ref} = (V_{a,b,c}^{P}-V_{s,i}).{U}_\mathit{output}+V_{ESS,i}^{ref}\\
\label{PRA}
\end{equation}
where $V_{ESS,i}^{ref}$ is the nominal voltage of the single-phase PV inverter. It is worth mentioning that $V_{a}^{P}$, $V_{b}^{P}$ and $V_{c}^{P}$ indicate phase voltage of 1st, 2nd and 3rd single-phase PV inverters connected to the three-phase PCC terminal respectively. Note that the reference output voltage of the ESS $i$ is also applied to $ith$ single-phase PV inverter as a nominal voltage. 

Finally, the reference currents ($i_{d,i}^{ref}$ and $i_{q,i}^{ref}$) in dq frame are calculated as:
\begin{align} 
\begin{split}
i_{d,i}^{ref}= 2(P_{ESS,i}^{ref})/V_{d,i}^{ref} \\
i_{q,i}^{ref}=-2(Q_{ESS,i}^{ref})/V_{d,i}^{ref}
\end{split}
\label{PRA5}
\end{align}

\begin{figure*}[t]
\begin{subfigure}{.48\textwidth}
  \centering
\includegraphics[width=\linewidth]{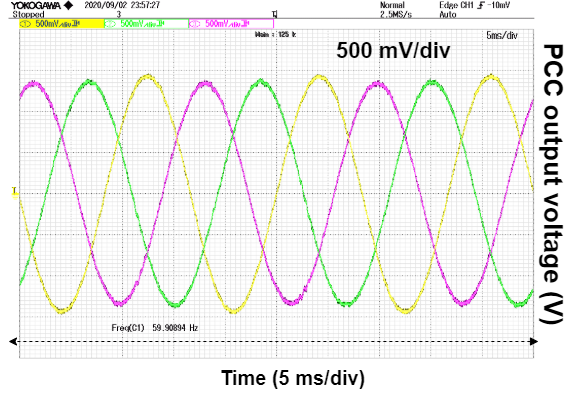} 
  \caption{}
\label{6a}
\end{subfigure}
\begin{subfigure}{.48\textwidth}
  \centering
  \includegraphics[width=\linewidth]{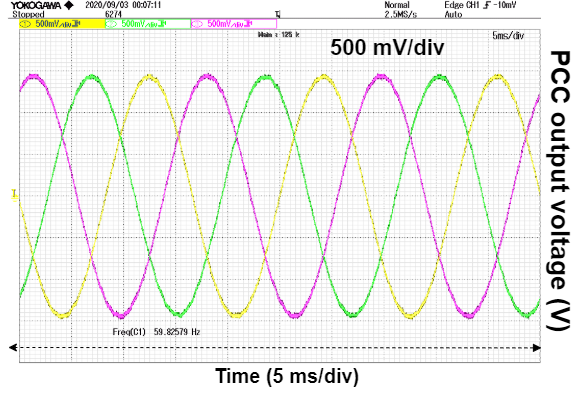}  
  \caption{}
\label{RPCA}
\end{subfigure}
\caption{Output voltage of PCC before (a) and (b) after implementing RPC.}
\label{RPCA}
\vspace{-8pt}
\end{figure*} 

\begin{figure}
\centering
\includegraphics[width=\linewidth]{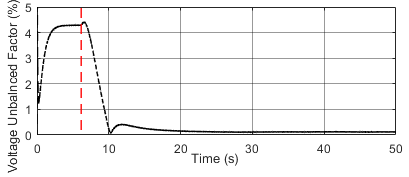}
\caption{VUF of PCC voltage.}
\label{VUF}
\vspace{-1pt}
\end{figure}

\begin{figure}
\centering
\includegraphics[width=\linewidth]{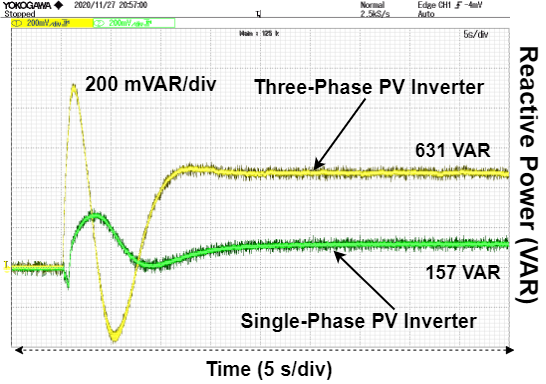}
\caption{Reactive power sharing performance between PV inverters.}
\label{RPSA}
\vspace{-15pt}
\end{figure}

\begin{figure*}[t]
\begin{subfigure}{.48\textwidth}
  \centering
\includegraphics[width=\linewidth]{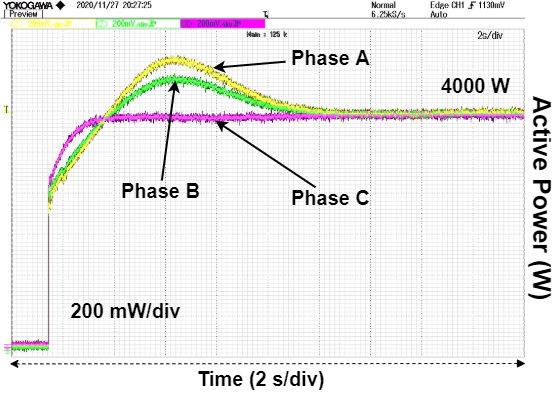} 
  \caption{}
\end{subfigure}
\begin{subfigure}{.48\textwidth}
  \centering
  \includegraphics[width=\linewidth]{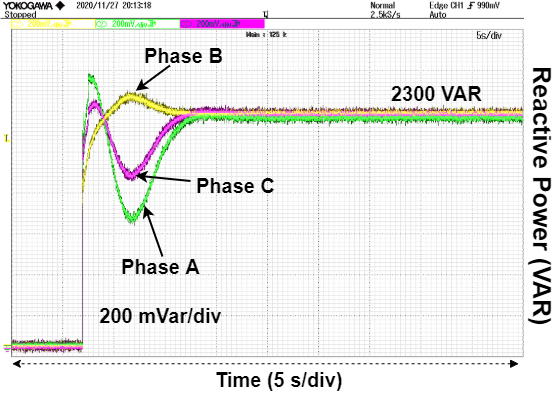}  
  \caption{}
\end{subfigure}
\caption{Active and reactive powers for phases A, B, and C at the three-phase PCC terminal}
\label{PRA}
\end{figure*} 
\vspace{1.5pt}
\subsection{PR current controller}
This controller regulates the output current ($i_{s,i}$) of the ESS $i$. The input to the PR controller is the reference current ($i_{i}^{ref}$) generated by single-phase sine generator \cite{single}. $i_{i}^{ref}$ is compared with $i_{s,i}$ and the error is received by the PR controller to generate the reference signals (duty cycle) to be fed to the pulse width modulator (PWM) block. Finally, the PWM controls the switching of the inverter based on this reference. More details are discussed in \cite{SOGI}. 

\section{real time simulation results}
This section shows the real-time simulation results based on the islanded operation of the multi-microgrid test system in Fig.~\ref{Simplified} to verify the proposed ESS multi-function controller in Fig.~\ref{Detailed} using  Opal-RT  OP5600  real-time  simulator in Fig.~\ref{opal}. The proposed ESS control is connected in parallel to the single-phase PV inverter and the single-phase load. The test system simulated is composed of a three-phase PV MG and a single-phase PV MG connected to phase line A of the three-phase MG. In addition, the power stage and control parameters are listed in Table~\ref{table}. 

\subsection{Performance of the RPC} Fig.~\ref{RPCA} depicts the PCC output voltage before and after compensation respectively. Fig.~\ref{RPCA}(a) shows the performance of the unbalanced MG, with no proposed RPC. The  output voltage at PCC terminal is unbalanced before compensation. It  can  be  seen  from Fig.~\ref{RPCA}(b) that voltage quality of the three-phase PCC is improved after implementing the proposed controller. In addition, the voltage unbalance factor (VUF) measuring the network unbalance at the PCC terminal demonstrates the effects of the proposed RPC. The values of $VUF$ is calculated as:  
\begin{equation}
\%VUF= \frac{\sqrt{(V_{dq}^{-1})^{2}}}{\sqrt{(V_{dq}^{+1})^{2}}}*100.
\label{percent}
\end{equation}
where $V_{dq}^{-1}$ and $V_{dq}^{+1}$ represent the magnitude of the negative and positive sequence of the fundamental voltage of $v_{pcc}^{3\phi}$ in $dq$ frame at PCC respectively. Fig.~\ref{VUF} shows the voltage unbalance factor (VUF) of $v_{pcc}^{3\phi}$ is reduced from 4.3\% to 0.03\% with the activation of the proposed controller at 7 second. 

\subsection{Performance of the RPSA} To verify the effectiveness of the proposed  RPSA method under the unbalanced MG and load conditions, the reactive power sharing performance of the three-phase and the single-phase PV inverter is presented in Fig.~\ref{RPSA}. Reactive power sharing proportionally between different ratings of three-phase and single-phase PV inverters is presented after implementing RPSA as shown in Fig.~\ref{RPSA}.


\begin{table}[t]
\caption{Simulation parameters}
\vspace{-2pt}
\begin{center}
\label{table}
\begin{tabular}{|c|c|}
\hline
\textbf{ESS System Parameter}   & \textbf{Value} \\ \hline
Switching frequency   & $F_{sw}$=20 kHz \\ \hline
LC filter  &     $L$=1.5 mH $C$=100 µF\\  \hline
DC link voltage & $v_{ESS,dc}^{ref}$=300 V\\  \hline
Nominal voltage    &  $V_{ESS}^{ref}$=120 rms V\\ \hline
System frequency  & $f_{ESS}^{ref}$=60 rad/s \\ \hline
\textbf{IA Control Parameter} & \textbf{RPC/ RPSA/ PBR  Value} \\ \hline
Integral     coefficient ($K_{i}$)                              &   10e-6, 10e-6, 10e-6 \\ \hline
Differential coefficient ($K_{d}$)                             &  5, 1, 2 \\ \hline
\textbf{PV and Load Parameter} & \textbf{ Power	Value} \\ \hline
PV Power &       $P_{PV}^{\phi}$ =3 kW, $P_{PV}^{3\phi}$=12 kW \\ \hline
$P_{L,i}^{\phi,ref}$, $Q_{L,i}^{\phi,ref}$   &    800 W, 500 VAR          \\ \hline
\textbf{PR Control Parameter} & \textbf{Coefficient Value} \\ \hline
$k_{pI}$, $w_{cI}$                           &     10,   1             \\\hline
$k_{rIk}$                           &   100(k = 1), 50(h = 3), 20(h = 5,7)           \\ \hline
\end{tabular}
\vspace{-15pt}
\end{center}
\end{table}

\subsection{Performance of the PBR} 
The effect of the proposed PBR strategy to balance each phase of the three-phase PCC power and eliminate unbalanced PCC voltage in unbalanced MGs is depicted in Fig.~\ref{PRA}. It can be seen from Fig.~\ref{PRA} that the three-phase active and reactive powers between phases A, B, and C of the three-phase at the PCC terminal are equal and therefore the three-phase PCC powers are balanced. 


\section{conclusion}
In this study, an energy storage system (ESS) using a novel multi-function control is proposed for voltage and power quality improvement and reactive power sharing in the islanded unbalanced multimicrogrids
(MMG). The multi-function control scheme consists of a reactive  power  compensation (RPC), a reactive power  sharing  algorithm (RPSA) and a power balance regulator (PBR). While the RPC and the PBR in ESS are utilized for reactive power compensation and balancing three-phase power of the PCC terminal respectively, the RPSA reduces reactive power sharing error among MMG by injecting or absorbing powers by the ESS. The incremental algorithm (IA) applied to the RPC, RPSA and PBR is developed and the algorithm inherently contains anti-windup to improves the stability of the system. The proposed control strategy has been tested experimentally in the real time simulator under the single-phase PV system and the unbalanced load connected to the three-phase PV system. The experimental results proved that the proposed control strategy improved the voltage and power unbalance and the reactive power sharing at the three-phase PCC.


%


\vspace{-4pt}
\bibliographystyle{IEEEtran}
\bibliography{IEEEexample}
\end{document}